%% file: ECOC_LaTeX_Template.tex
\documentclass[a4paper, oneside, twocolumn, notitlepage, 10pt]{extarticle_ecoc}
\usepackage{ecoc}

\usepackage{amsmath,amssymb,amsfonts}
\usepackage{algorithmic}
\usepackage{graphicx}
\usepackage{textcomp}
\usepackage{xcolor}
\usepackage{orcidlink}

\usepackage{tikz}
\usepackage{pgfplots}
\usepackage{mathtools}
\pgfplotsset{compat=1.18}
\usepgfplotslibrary{groupplots}
\usepgfplotslibrary{colormaps}
\usepgfplotslibrary{colorbrewer}
\usepgfplotslibrary{external} 
\usetikzlibrary{arrows.meta}
\usetikzlibrary{arrows}

\definecolor{uclwhite}{RGB}{250, 250, 250}
\definecolor{ucldpurple}{RGB}{54, 26, 84}    %
\definecolor{uclbpurple}{RGB}{153, 59, 255}  %
\definecolor{uclmpurple}{RGB}{186, 130, 255} %
\definecolor{ucllpurple}{RGB}{221, 189, 255} %
\definecolor{uclppurple}{RGB}{238, 222, 255} %
\definecolor{uclhblue}{RGB}{48, 214, 255}    %

\definecolor{uclg0}{RGB}{54,26,84}    %
\definecolor{uclg1}{RGB}{153,59,255}  %
\definecolor{uclg2}{RGB}{0,46,166}    %
\definecolor{uclg3}{RGB}{120,28,28}   %
\definecolor{uclg4}{RGB}{158,26,84}   %
\definecolor{uclg5}{RGB}{0,94,92}     %
\definecolor{uclg6}{RGB}{84,135,255}  %
\definecolor{uclg7}{RGB}{245,99,0}    %
\definecolor{uclg8}{RGB}{237,54,125}  %
\definecolor{uclg9}{RGB}{0,158,156}   %

\definecolor{o-band}{HTML}{a8d6f4}
\definecolor{e-band}{HTML}{ffbbd5}
\definecolor{s-band}{HTML}{c6afe9}
\definecolor{c-band}{HTML}{8bdebf}
\definecolor{l-band}{HTML}{ffcaa6}

\begin{document}
\selectlanguage{english}    %

\title{GPU-Accelerated Optimisation of Symmetric Bidirectional Ultrawideband Coherent
Transmission Under Launch Power Constraints}%

\author{
    Mindaugas~Jarmolovi\v{c}ius\textsuperscript{(1)},
    Eric~Sillekens\textsuperscript{(1)},
    Polina~Bayvel\textsuperscript{(1)},
    Robert~I.~Killey\textsuperscript{(1)}
}

\maketitle                  %

\begin{strip}
    \begin{author_descr}

        \textsuperscript{(1)} 
        Optical Networks Group, UCL (University College London), London, UK,
        \textcolor{blue}{\uline{zceemja@ucl.ac.uk}}
        
    \end{author_descr}
\end{strip}

\renewcommand\footnotemark{}
\renewcommand\footnoterule{}

\begin{strip}
    \begin{ecoc_abstract}
        We optimise bidirectional OESCL-band coherent transmission under total fibre power constraints using a GPU-accelerated boundary-value Raman solver and launch-power optimisation. While no gain is observed without power constraints, for an 18-dBm limit 3-span transmission reversing the O-band direction increases aggregate capacity by up to 24.5\%. \copyright2026 The Author(s)
    \end{ecoc_abstract}
\end{strip}

\section{Introduction}

Coherent optical transmission at wavelengths beyond the C-band has emerged as an attractive route to increasing link capacity in areas where new fibre installation is not cost-effective \cite{luis_450_2026,shimizu_186-thz_2025}. Recent advances in amplification technology in O-, E-, S-bands \cite{donodinDopedFiberAmplifiers2025} have pushed datarates in ultra-wideband (UWB) experiments using standard single-mode fibres (SSMF) to as high as 450~Tb/s in deployed fibre links \cite{luis_450_2026} utilising OESCL-bands, demonstrating the potential of future UWB systems.

\begin{figure}[b]
\vspace{-1.5em}
    \begin{center}
    \includegraphics[width=\linewidth]{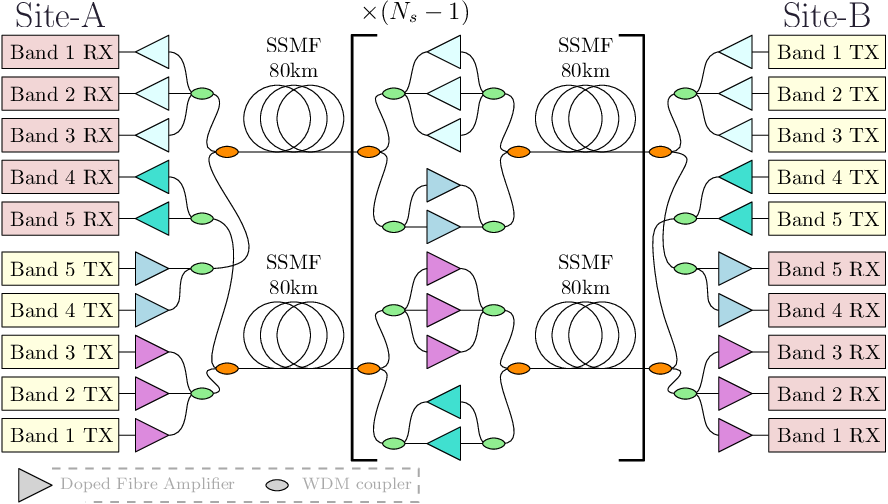}
    \caption{Diagram of a bidirectional transmission setup with an example forward/backward split of two/three bands. }
    \label{fig:setup}
    \end{center}
\end{figure}

However, deployed systems suffer from optical power limitations due to safety regulations. Previous analyses show that the total link throughput decreases with more strict power limits \cite{jarmoloviciusUltrawidebandOpticalFibre2024,yangChallengesBreakthroughsModeling2026}, as channels are launched with sub-optimum power. Assuming that deployed links will commonly have a symmetric fibre pair to transmit data in both directions, such power-limited system can be improved by transmitting some bands backwards, as this allows an increase in the per-band launch power, without exceeding the total power constraints at any point along the fibre. This has been shown to achieve higher datarates in deployed fibre links \cite{yang1355TbOBand} showing that bidirectional band allocation can improve total throughput under fibre power limits, but practical optimisation over forward/backward band permutations requires repeated solution of Raman power evolution with counter-propagating bands. 

Here we explore the ultra-wideband optimisation framework to such bidirectional scenarios and make it computationally practical using a GPU-accelerated boundary-value solver. We extend the existing models to enable optimisation of bidirectional band systems and evaluate performance for a number of scenarios for OESCL-band transmission.

\section{Bidirectional Transmission Modelling}

The transmission link diagram is shown in Figure~\ref{fig:setup}, assuming that the uplink and downlink between site-A and site-B are symmetric. 
To evaluate the performance of a bidirectional system, we define SNR as follows:
\begin{equation}
\mathrm{SNR}_i \approx \frac{P_i}{N_{\rm s} P_{\mathrm{ASE}_{i}} + \eta_{\mathrm{NLI},i}P_i^3 + P_i \kappa_i},
\label{eq:snr}
\end{equation}

Where $i$ is the channel index, $P_i$ is the channel launch power, $N_{\rm s}$ is the number of spans, $\mathrm{ASE}_{i}$ is the amplified spontaneous emission (ASE) noise from the optical amplifiers. ASE noise arising from the ISRS power transfer is included, however it is negligible for all scenarios. $\eta_\mathrm{NLI}$ is non-linear interference (NLI) noise power which is calculated using a fast Gaussian noise integral model \cite{jarmoloviciusOptimisingOtoUBand2024b}. NLI interaction between co-propagating signals is assumed to be negligible, as propagating signals do not overlap within the same wavelength. $\kappa_i$ is channel transceiver noise (obtained from back to back SNR), which is assumed to be 20~dB throughout this work. WDM coupler loss is assumed to be negligible.

Two  80~km fibres links are investigated -- G.652.D-compliant fibre profile from a deployed system (Fibre-A) \cite{sohanpalUltraWidebandTransmissionSystems2026,luis_450_2026} and modelled low loss fibre (Fibre-B) \cite{sohanpalUltraWidebandTransmissionSystems2026}. %
Frequency dependent nonlinear coefficient, effective area and Raman spectrum gain parameters used are described in our earlier work \cite{jarmoloviciusUltrawidebandOpticalFibre2024}. Transmission channels are 150~GHz spaced, with 140~GBaud dual-polarisation Gaussian-shaped constellations. Parameters of each band are shown in Table~\ref{tab:band_params}.

\begin{table}
    \centering
    \begin{tabular}{c|c|c|c}
        Band & Wavelengths & $N_\mathrm{ch}$ & NF \\\hline
        L & 1567--1624~nm & 43 & 6~dB \\
        C & 1530--1565~nm & 28 & 5~dB \\
        S & 1460--1527~nm & 59 & 7~dB \\
        E & 1400--1455~nm & 53 & 6.5~dB \\
        O & 1260--1360~nm & 116 & 5~dB \\
    \end{tabular}
    \caption{Band parameters, NF representing amplifier noise figure, $N_\mathrm{ch}$ number of channels in a band}
    \label{tab:band_params}
\end{table}

For optimisation scenarios with power limits, the optimiser must consider the nonlinear boundary condition constraint as defined in Eq.~\ref{eq:nlconstrain}:
\vspace{-.8em}
\begin{equation}\label{eq:nlconstrain}
    -\infty < \log(\max\left(\sum_{\forall i} P_{i,z}\right)) \leq \log(P_\mathrm{max})
\end{equation}

where $z$ is the distance along the fibre, $P_\mathrm{max}$ is the target power limit, and $P_{i,z}$ is the power evolution obtained from Raman coupled ordinary differential equations (ODEs). The constrained function is transformed with logarithm for consistency. COBYQA optimiser \cite{rago_thesis,razh_cobyqa} was used as it showed excellent performance in this particular set of scenarios compared to similar nonlinear constrained problem optimisers in terms of stability and the required number of loss function evaluations to obtain the optimum launch power. Using COBYQA optimiser requires rapid power evolution solver as it requires a large number of nonlinear constraint evaluations for each optimisation iteration. This can become prohibitively expensive to compute when bands are transmitted backwards, as power evaluation solver changes from an initial value problem (IVP) for a system of ODEs where power at the start of the fibre is known, to a boundary value problem (BVP) where power at the start of the fibre for backward transmitted bands is not known. To tackle this problem, firstly, variable transformation is implemented, as proposed in \cite{hafermannFastStableMethod2025}. Secondly, a GPU accelerated BVP ODE solver is implemented\footnote{Source\,\,code:~\url{https://gist.github.com/RicardoDominguez/f013d21a5991e863ffcf9076f5b9b34d}} based on \cite{dessoleMassivelyParallelAlgorithm2021a}, using 3-point Lobatto quadrature for residual estimation. This algorithm achieves solution  relative error of $< 10^{-10}$ within one iteration of backtracking step, which is sufficient to correctly estimate optical power at the start of the fibre to satisfy backward power conditions. This achieves power evolution computation times in the range of 20~ms~--~100~ms on a single Nvidia V100 GPU, depending on the number of backward bands.

\begin{figure*}[t]
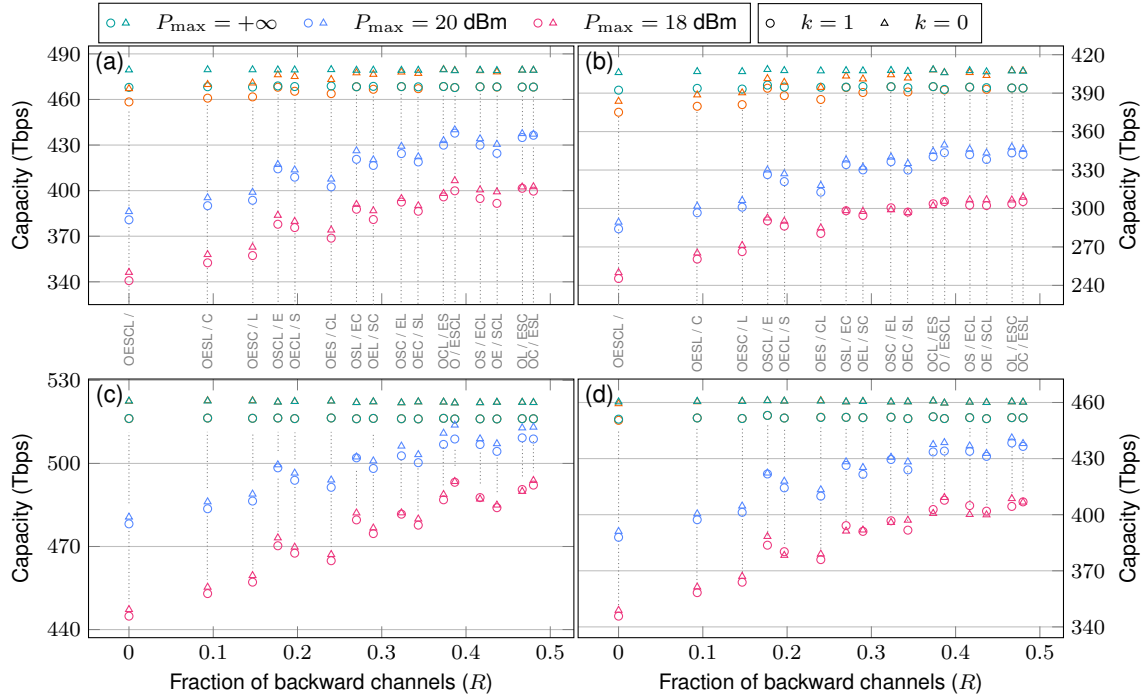

    \begin{center}
    \include{figure}
    \caption{Aggregate capacity per fibre link of scenarios with some bands being send backwards, normalised to a number of total number of channels. Scenario (a) is Fibre-A 1$\times$80~km link, (b) Fibre-A 3$\times$80~km link, (c) Fibre-B 1$\times$80~km link, (d) Fibre-B 3$\times$80~km link. }
    \label{fig:result}
    \end{center}
\end{figure*}

\section{Results}
All possible permutations of forward-backward band combinations were evaluated for the 80~km link, as shown in Figure~\ref{fig:result}a and ~\ref{fig:result}c, and 3$\times$80~km link shown in Figure~\ref{fig:result}b and ~\ref{fig:result}d for two fibre scenarios. For these links, the optimisation is performed with three different power constraints, with a maximum power of 18~dBm, 20~dBm and no power limit. For each of these scenarios, we use two different loss function coefficients -- $k=0$ and $k=1$. $k$ is SNR flatness weight as shown in Eq.~\ref{eq:loss}, which forces the optimiser to achieve a more uniform transmission quality within each band. $\overline{X}$ denotes the mean value of vector $X$. %
\vspace{-8pt}
\begin{equation}\label{eq:loss}
    \mathcal{L} = \sum_{b \in \{O,E,S,C,L\}} \overline{\log(\mathrm{SNR}_b)} - \mathrm{std}(\log(\mathrm{SNR}_b))\cdot k
\end{equation}

In this work the aggregate capacity was evaluated using the capacity limit $C=\sum_i 2 B (1 + \log_2(\mathrm{SNR_i}))$, where $B$ is symbol rate and $\mathrm{SNR_i}$ includes all channels $i$ within a single fibre link regardless of the propagation direction. %
COBYQA reached optimum with 90--550 loss function evaluations (including expensive NLI evaluations) and 2000--13000 nonlinear constraint evaluations.

The results show that, in each scenario with no power constraints, the benefit of separating bands into forward and backward directions is minimal, with aggregate capacity best improvements of 0.23\% in the 1-span scenarios and below 1\% in the 3-span scenarios. 

\begin{figure}[t]
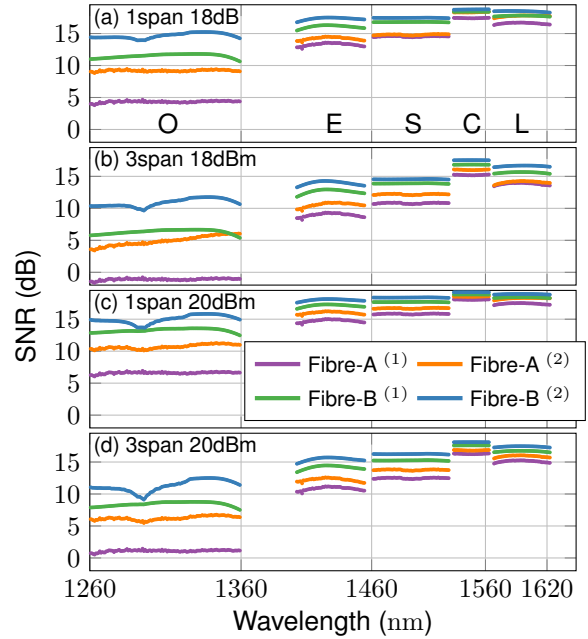

    \begin{center}
    \include{snr}
    \vspace{-2em}
    \caption{SNR results (with $k=1$) comparing no-split $^{(1)}$ versus O-Band only propagating in opposite direction $^{(2)}$.}
    \label{fig:snr_results}
    \end{center}
    \vspace{-1em}
\end{figure}

We define $R$ as a fraction of backward propagating channels normalised to a number of total number of channels. Results with power constraints show as $R$ approaches $0.5$, the aggregate capacity increases as optical power gets distributed at the edges of fibre and channel powers can be increased to reach values closer to the optimum. However, due to the attenuation profile, ISRS effects and higher nonlinearities in the O-band due to zero-dispersion, certain forward/backward split combinations achieve better performance than scenarios with forward/backward cases with $R$ closer to 0.5. Most notably, E/OSCL split compared to CL/OES and OES/CL split, and O/ESCL compared to the other higher split ratios, stand out as having better performance. O/ESCL, OL/ESC and OC/ESL scenarios have the best performance, with the largest difference between these three combinations being below 1\%. Therefore, in power-constrained scenarios an O/ESCL would be preferable due to OL/ESC and OC/ESL requiring an unconventional WDM coupler setup. The SNR results showing improvements of O/ESCL split are shown in Fig.~\ref{fig:snr_results} demonstrating that most SNR improvements are in the O-band.

The general improvements for the separation of bands versus non-splitting are shown in Table~\ref{tab:nolim_results}. In multi-span setups the capacity gain is less dependent on band slitting as achievable O-Band performance becomes limited, as shown by low O-Band SNR values in Fig.~\ref{fig:snr_results}.

\bgroup
\def\arraystretch{1}%
\setlength\tabcolsep{4pt}%
\begin{table}[t]
    \centering
    \begin{tabular}{c|c|c}
        \hline

        Fibre-A & $P_\mathrm{max}=$20~dBm & $P_\mathrm{max}=$18~dBm \\\hline
        1~span & +57.05 (+15.0\%) & +60.80 (+17.8\%) \\
        3~span & +59.41 (+20.9\%) & +60.03 (+24.5\%) \\\hline
        \hline
        Fibre-B & $P_\mathrm{max}=$20~dBm & $P_\mathrm{max}=$18~dBm \\\hline
        1~span & +31.05 (+6.5\%) & +48.33 (+10.9\%) \\
        3~span & +50.45 (+13.0\%) &+61.96 (+17.9\%) \\
        
    \end{tabular}
    \caption{Best aggregate capacity improvements (in Tbps) compared to no band splitting scenario for $k=1$.}
    \label{tab:nolim_results}
\end{table}
\egroup

\section{Conclusion}

In this work, the solver for coupled Raman equations for BVP ODEs are implemented for a fast evaluation on a GPU. COBYQA optimiser is used to quantify UBW transmission performance. The proposed optimisation scheme is versatile and could be used for other nonlinear constraint problems such as distributed Raman amplification optimisation. We explore combinations of band transmission in both directions along single mode fibre links with optical power limit as the nonlinear constraint, estimating aggregate capacity gain. Results demonstrate that with a 20~dBm power constraint, aggregate single span link capacity can be increased by 15\% when transmitting only O-band in the opposite direction to ESCL-bands, followed by O-band together with one other band. The least gain case is OE-Band combination being transmitted in the same direction. This shows that for best performance, band placement must be such that the O-band has the lowest penalty caused by ISRS power transfer. Results further show that improvements are higher in multi-span setups, in scenarios with more strict power constraints, and in scenario involving higher attenuation fibres. The highest aggregate capacity improvements are shown to be 24.5\% in 3-span, 18~dBm power limit deployed fibre setup.

\clearpage
\section{Acknowledgements}
This work was supported by EPSRC grants EP/R035342/1 TRANSNET and EP/W015714/1 EWOC.
Professor Dame Polina Bayvel is supported by a Royal Society Research Professorship and Dr Eric Sillekens by Department for Science, 
Innovation and Technology and the Royal Academy of Engineering under the Research Fellowship scheme. \\ The authors thank Dr~Ronit~Sohanpal for valuable discussions.

\printbibliography %

\vspace{-4mm}

\end{document}

%% file: figure.tex
    \begin{tikzpicture}[baseline]\footnotesize

\pgfplotsset{dual-line/.style 2 args={legend image code/.code={
      \draw[#1] plot coordinates { (-.2cm,0cm) };%
      \draw[#2] plot coordinates { (0cm,0cm) };%
}}}

    \begin{groupplot}[
        group style={
            group name=power_graph,
            group size=2 by 2,
            x descriptions at=edge bottom,
            y descriptions at=all,
            horizontal sep=.2em,
            vertical sep=3.5em,
        },
        width=.5\linewidth,
        height=5cm,
        ylabel={Capacity (Tbps)},
        xlabel={Fraction of backward channels ($R$)},
        ymajorgrids=true,
        xmajorgrids=false,
        clip=false,
    ]

    \nextgroupplot[
        legend style={font=\footnotesize,at={(rel axis cs:0.02,1.01)},legend columns=3,anchor=south west,column sep=1em},
        legend entries={$P_\mathrm{max}=+\infty$,$P_\mathrm{max}=20~\text{dBm}$, $P_\mathrm{max}=18~\text{dBm}$},
        ytick={310,340,...,550},
        ymin=325
    ]

    \addlegendimage{dual-line={uclg9,only marks,mark=o,mark size=1.5pt}{uclg9,only marks,mark=triangle,mark size=1.5pt}};
    \addlegendimage{dual-line={uclg6,only marks,mark=o,mark size=1.5pt}{uclg6,only marks,mark=triangle,mark size=1.5pt}};
    \addlegendimage{dual-line={uclg8,only marks,mark=o,mark size=1.5pt}{uclg8,only marks,mark=triangle,mark size=1.5pt}};
    
    \addplot[uclg7,only marks,mark=o,mark size=1.5pt] table[x=num,y=cap]{data/NDFF1spans80kmp25_fw_bw_signals7b.txt};
    \addplot[uclg6,only marks,mark=o,mark size=1.5pt] table[x=num,y=cap]{data/NDFF1spans80kmp20_fw_bw_signals7b.txt};
    \addplot[uclg8,only marks,mark=o,mark size=1.5pt] table[x=num,y=cap]{data/NDFF1spans80kmp18_fw_bw_signals7b.txt};
    \addplot[uclg9,only marks,mark=o,mark size=1.5pt] table[x=num,y=cap]{data/NDFF1spans80kmpnan_fw_bw_signals7b.txt};

    \addplot[uclg7,only marks,mark=triangle,mark size=1.5pt] table[x=num,y=cap]{data/NDFF1spans80kmp25_fw_bw_signals8b.txt};
    \addplot[uclg6,only marks,mark=triangle,mark size=1.5pt] table[x=num,y=cap]{data/NDFF1spans80kmp20_fw_bw_signals8b.txt};
    \addplot[uclg8,only marks,mark=triangle,mark size=1.5pt] table[x=num,y=cap]{data/NDFF1spans80kmp18_fw_bw_signals8b.txt};
    \addplot[uclg9,only marks,mark=triangle,mark size=1.5pt] table[x=num,y=cap]{data/NDFF1spans80kmpnan_fw_bw_signals8b.txt};

\draw[gray,densely dotted] (axis cs:0.000,390.1) -- (axis cs:0.000,453.8);
\draw[gray,densely dotted] (axis cs:0.147,402.9) -- (axis cs:0.147,457.1);
\draw[gray,densely dotted] (axis cs:0.197,417.8) -- (axis cs:0.197,460.7);
\draw[gray,densely dotted] (axis cs:0.343,426.5) -- (axis cs:0.343,462.4);
\draw[gray,densely dotted] (axis cs:0.177,421.4) -- (axis cs:0.177,463.4);
\draw[gray,densely dotted] (axis cs:0.323,433.5) -- (axis cs:0.323,463.8);
\draw[gray,densely dotted] (axis cs:0.373,437.3) -- (axis cs:0.373,463.8);
\draw[gray,densely dotted] (axis cs:0.480,441.5) -- (axis cs:0.480,463.5);
\draw[gray,densely dotted] (axis cs:0.387,444.5) -- (axis cs:0.387,463.1);
\draw[gray,densely dotted] (axis cs:0.467,441.9) -- (axis cs:0.467,463.5);
\draw[gray,densely dotted] (axis cs:0.417,438.5) -- (axis cs:0.417,463.7);
\draw[gray,densely dotted] (axis cs:0.270,430.5) -- (axis cs:0.270,463.7);
\draw[gray,densely dotted] (axis cs:0.437,434.8) -- (axis cs:0.437,463.5);
\draw[gray,densely dotted] (axis cs:0.290,424.5) -- (axis cs:0.290,462.0);
\draw[gray,densely dotted] (axis cs:0.240,411.7) -- (axis cs:0.240,459.2);
\draw[gray,densely dotted] (axis cs:0.093,399.3) -- (axis cs:0.093,456.3);

\draw[gray,densely dotted] (axis cs:0.000,337.4) -- (axis cs:0.000,325) node [rotate=90,left=3pt,fill=white,font=\tiny,inner sep=0pt]{OESCL / };
\draw[gray,densely dotted] (axis cs:0.147,353.7) -- (axis cs:0.147,325) node [rotate=90,left=3pt,fill=white,font=\tiny,inner sep=0pt]{OESC / L};
\draw[gray,densely dotted] (axis cs:0.197,371.9) -- (axis cs:0.197,325) node [rotate=90,left=3pt,fill=white,font=\tiny,inner sep=0pt]{OECL / S};
\draw[gray,densely dotted] (axis cs:0.343,382.6) -- (axis cs:0.343,325) node [rotate=90,left=3pt,fill=white,font=\tiny,inner sep=0pt]{OEC / SL};
\draw[gray,densely dotted] (axis cs:0.177,374.1) -- (axis cs:0.177,325) node [rotate=90,left=3pt,fill=white,font=\tiny,inner sep=0pt]{OSCL / E};
\draw[gray,densely dotted] (axis cs:0.323,388.5) -- (axis cs:0.323,325) node [rotate=90,left=3pt,fill=white,font=\tiny,inner sep=0pt]{OSC / EL};
\draw[gray,densely dotted] (axis cs:0.373,391.9) -- (axis cs:0.373,325) node [rotate=90,left=3pt,fill=white,font=\tiny,inner sep=0pt]{OCL / ES};
\draw[gray,densely dotted] (axis cs:0.480,395.7) -- (axis cs:0.480,325) node [rotate=90,left=3pt,fill=white,font=\tiny,inner sep=0pt]{OC / ESL};
\draw[gray,densely dotted] (axis cs:0.387,395.9) -- (axis cs:0.387,325) node [rotate=90,left=3pt,fill=white,font=\tiny,inner sep=0pt]{O / ESCL};
\draw[gray,densely dotted] (axis cs:0.467,397.5) -- (axis cs:0.467,325) node [rotate=90,left=3pt,fill=white,font=\tiny,inner sep=0pt]{OL / ESC};
\draw[gray,densely dotted] (axis cs:0.417,390.9) -- (axis cs:0.417,325) node [rotate=90,left=3pt,fill=white,font=\tiny,inner sep=0pt]{OS / ECL};
\draw[gray,densely dotted] (axis cs:0.270,383.9) -- (axis cs:0.270,325) node [rotate=90,left=3pt,fill=white,font=\tiny,inner sep=0pt]{OSL / EC};
\draw[gray,densely dotted] (axis cs:0.437,387.7) -- (axis cs:0.437,325) node [rotate=90,left=3pt,fill=white,font=\tiny,inner sep=0pt]{OE / SCL};
\draw[gray,densely dotted] (axis cs:0.290,377.2) -- (axis cs:0.290,325) node [rotate=90,left=3pt,fill=white,font=\tiny,inner sep=0pt]{OEL / SC};
\draw[gray,densely dotted] (axis cs:0.240,365.1) -- (axis cs:0.240,325) node [rotate=90,left=3pt,fill=white,font=\tiny,inner sep=0pt]{OES / CL};
\draw[gray,densely dotted] (axis cs:0.093,349.0) -- (axis cs:0.093,325) node [rotate=90,left=3pt,fill=white,font=\tiny,inner sep=0pt]{OESL / C};

\draw[gray,densely dotted] (axis cs:0.000,378.5) -- (axis cs:0.000,347.6);
\draw[gray,densely dotted] (axis cs:0.147,391.0) -- (axis cs:0.147,364.4);
\draw[gray,densely dotted] (axis cs:0.197,405.4) -- (axis cs:0.197,383.2);
\draw[gray,densely dotted] (axis cs:0.343,413.9) -- (axis cs:0.343,394.2);
\draw[gray,densely dotted] (axis cs:0.177,408.9) -- (axis cs:0.177,385.4);
\draw[gray,densely dotted] (axis cs:0.323,420.6) -- (axis cs:0.323,400.3);
\draw[gray,densely dotted] (axis cs:0.373,424.3) -- (axis cs:0.373,403.8);
\draw[gray,densely dotted] (axis cs:0.480,428.4) -- (axis cs:0.480,407.7);
\draw[gray,densely dotted] (axis cs:0.387,431.3) -- (axis cs:0.387,407.9);
\draw[gray,densely dotted] (axis cs:0.467,428.8) -- (axis cs:0.467,409.6);
\draw[gray,densely dotted] (axis cs:0.417,425.5) -- (axis cs:0.417,402.7);
\draw[gray,densely dotted] (axis cs:0.270,417.7) -- (axis cs:0.270,395.5);
\draw[gray,densely dotted] (axis cs:0.437,421.9) -- (axis cs:0.437,399.5);
\draw[gray,densely dotted] (axis cs:0.290,411.9) -- (axis cs:0.290,388.6);
\draw[gray,densely dotted] (axis cs:0.240,399.5) -- (axis cs:0.240,376.1);
\draw[gray,densely dotted] (axis cs:0.093,387.4) -- (axis cs:0.093,359.5);

    \nextgroupplot[
    ytick pos=right, ylabel near ticks,
    legend style={font=\footnotesize,at={(rel axis cs:0.37,1.01)},legend columns=2,anchor=south west,column sep=1em},
    legend entries={$k=1$,$k=0$},
    ytick={210,240,...,450},
    ymin=225
    ]

    \addlegendimage{black,only marks,mark=o,mark size=1.5pt};
    \addlegendimage{only marks,mark=triangle,mark size=1.5pt};

    \addplot[uclg7,only marks,mark=o,mark size=1.5pt] table[x=num,y=cap]{data/NDFF3spans80kmp25_fw_bw_signals7b.txt};
    \addplot[uclg6,only marks,mark=o,mark size=1.5pt] table[x=num,y=cap]{data/NDFF3spans80kmp20_fw_bw_signals7b.txt};
    \addplot[uclg8,only marks,mark=o,mark size=1.5pt] table[x=num,y=cap]{data/NDFF3spans80kmp18_fw_bw_signals7b.txt};
    \addplot[uclg9,only marks,mark=o,mark size=1.5pt] table[x=num,y=cap]{data/NDFF3spans80kmpnan_fw_bw_signals7b.txt};

    \addplot[uclg7,only marks,mark=triangle,mark size=1.5pt] table[x=num,y=cap]{data/NDFF3spans80kmp25_fw_bw_signals8b.txt};
    \addplot[uclg6,only marks,mark=triangle,mark size=1.5pt] table[x=num,y=cap]{data/NDFF3spans80kmp20_fw_bw_signals8b.txt};
    \addplot[uclg8,only marks,mark=triangle,mark size=1.5pt] table[x=num,y=cap]{data/NDFF3spans80kmp18_fw_bw_signals8b.txt};
    \addplot[uclg9,only marks,mark=triangle,mark size=1.5pt] table[x=num,y=cap]{data/NDFF3spans80kmpnan_fw_bw_signals8b.txt};
\draw[gray,densely dotted] (axis cs:0.000,292.2) -- (axis cs:0.000,371.4);
\draw[gray,densely dotted] (axis cs:0.147,309.3) -- (axis cs:0.147,377.2);
\draw[gray,densely dotted] (axis cs:0.197,330.4) -- (axis cs:0.197,384.2);
\draw[gray,densely dotted] (axis cs:0.343,338.4) -- (axis cs:0.343,387.0);
\draw[gray,densely dotted] (axis cs:0.177,333.3) -- (axis cs:0.177,389.9);
\draw[gray,densely dotted] (axis cs:0.323,343.5) -- (axis cs:0.323,391.1);
\draw[gray,densely dotted] (axis cs:0.373,347.9) -- (axis cs:0.373,391.1);
\draw[gray,densely dotted] (axis cs:0.480,349.6) -- (axis cs:0.480,389.9);
\draw[gray,densely dotted] (axis cs:0.387,353.1) -- (axis cs:0.387,388.5);
\draw[gray,densely dotted] (axis cs:0.467,351.4) -- (axis cs:0.467,390.1);
\draw[gray,densely dotted] (axis cs:0.417,349.8) -- (axis cs:0.417,390.7);
\draw[gray,densely dotted] (axis cs:0.270,341.2) -- (axis cs:0.270,390.5);
\draw[gray,densely dotted] (axis cs:0.437,346.8) -- (axis cs:0.437,389.1);
\draw[gray,densely dotted] (axis cs:0.290,335.4) -- (axis cs:0.290,386.6);
\draw[gray,densely dotted] (axis cs:0.240,321.1) -- (axis cs:0.240,381.2);
\draw[gray,densely dotted] (axis cs:0.093,304.8) -- (axis cs:0.093,376.0);

\draw[gray,densely dotted] (axis cs:0.000,242.9) -- (axis cs:0.000,225) node [rotate=90,left=4pt,fill=white,font=\tiny,inner sep=0pt]{OESCL / };
\draw[gray,densely dotted] (axis cs:0.147,263.7) -- (axis cs:0.147,225) node [rotate=90,left=4pt,fill=white,font=\tiny,inner sep=0pt]{OESC / L};
\draw[gray,densely dotted] (axis cs:0.197,283.4) -- (axis cs:0.197,225) node [rotate=90,left=4pt,fill=white,font=\tiny,inner sep=0pt]{OECL / S};
\draw[gray,densely dotted] (axis cs:0.343,293.9) -- (axis cs:0.343,225) node [rotate=90,left=4pt,fill=white,font=\tiny,inner sep=0pt]{OEC / SL};
\draw[gray,densely dotted] (axis cs:0.177,287.5) -- (axis cs:0.177,225) node [rotate=90,left=4pt,fill=white,font=\tiny,inner sep=0pt]{OSCL / E};
\draw[gray,densely dotted] (axis cs:0.323,296.0) -- (axis cs:0.323,225) node [rotate=90,left=4pt,fill=white,font=\tiny,inner sep=0pt]{OSC / EL};
\draw[gray,densely dotted] (axis cs:0.373,299.2) -- (axis cs:0.373,225) node [rotate=90,left=4pt,fill=white,font=\tiny,inner sep=0pt]{OCL / ES};
\draw[gray,densely dotted] (axis cs:0.480,302.3) -- (axis cs:0.480,225) node [rotate=90,left=4pt,fill=white,font=\tiny,inner sep=0pt]{OC / ESL};
\draw[gray,densely dotted] (axis cs:0.387,302.3) -- (axis cs:0.387,225) node [rotate=90,left=4pt,fill=white,font=\tiny,inner sep=0pt]{O / ESCL};
\draw[gray,densely dotted] (axis cs:0.467,300.4) -- (axis cs:0.467,225) node [rotate=90,left=4pt,fill=white,font=\tiny,inner sep=0pt]{OL / ESC};
\draw[gray,densely dotted] (axis cs:0.417,299.5) -- (axis cs:0.417,225) node [rotate=90,left=4pt,fill=white,font=\tiny,inner sep=0pt]{OS / ECL};
\draw[gray,densely dotted] (axis cs:0.270,295.3) -- (axis cs:0.270,225) node [rotate=90,left=4pt,fill=white,font=\tiny,inner sep=0pt]{OSL / EC};
\draw[gray,densely dotted] (axis cs:0.437,299.3) -- (axis cs:0.437,225) node [rotate=90,left=4pt,fill=white,font=\tiny,inner sep=0pt]{OE / SCL};
\draw[gray,densely dotted] (axis cs:0.290,291.6) -- (axis cs:0.290,225) node [rotate=90,left=4pt,fill=white,font=\tiny,inner sep=0pt]{OEL / SC};
\draw[gray,densely dotted] (axis cs:0.240,277.7) -- (axis cs:0.240,225) node [rotate=90,left=4pt,fill=white,font=\tiny,inner sep=0pt]{OES / CL};
\draw[gray,densely dotted] (axis cs:0.093,257.9) -- (axis cs:0.093,225) node [rotate=90,left=4pt,fill=white,font=\tiny,inner sep=0pt]{OESL / C};

\draw[gray,densely dotted] (axis cs:0.000,283.6) -- (axis cs:0.000,250.3);
\draw[gray,densely dotted] (axis cs:0.147,300.1) -- (axis cs:0.147,271.7);
\draw[gray,densely dotted] (axis cs:0.197,320.5) -- (axis cs:0.197,292.0);
\draw[gray,densely dotted] (axis cs:0.343,328.4) -- (axis cs:0.343,302.8);
\draw[gray,densely dotted] (axis cs:0.177,323.4) -- (axis cs:0.177,296.2);
\draw[gray,densely dotted] (axis cs:0.323,333.3) -- (axis cs:0.323,305.0);
\draw[gray,densely dotted] (axis cs:0.373,337.6) -- (axis cs:0.373,308.2);
\draw[gray,densely dotted] (axis cs:0.480,339.2) -- (axis cs:0.480,311.5);
\draw[gray,densely dotted] (axis cs:0.387,342.6) -- (axis cs:0.387,311.5);
\draw[gray,densely dotted] (axis cs:0.467,341.0) -- (axis cs:0.467,309.5);
\draw[gray,densely dotted] (axis cs:0.417,339.4) -- (axis cs:0.417,308.6);
\draw[gray,densely dotted] (axis cs:0.270,331.1) -- (axis cs:0.270,304.2);
\draw[gray,densely dotted] (axis cs:0.437,336.5) -- (axis cs:0.437,308.4);
\draw[gray,densely dotted] (axis cs:0.290,325.5) -- (axis cs:0.290,300.4);
\draw[gray,densely dotted] (axis cs:0.240,311.6) -- (axis cs:0.240,286.1);
\draw[gray,densely dotted] (axis cs:0.093,295.7) -- (axis cs:0.093,265.7);

    \nextgroupplot[
        ytick={410,440,...,550},
    ]

    \addplot[uclg7,only marks,mark=o,mark size=1.5pt] table[x=num,y=cap]{data/ULL-low-wp1spans80kmp25_fw_bw_signals7c.txt};
    \addplot[uclg6,only marks,mark=o,mark size=1.5pt] table[x=num,y=cap]{data/ULL-low-wp1spans80kmp20_fw_bw_signals7c.txt};
    \addplot[uclg8,only marks,mark=o,mark size=1.5pt] table[x=num,y=cap]{data/ULL-low-wp1spans80kmp18_fw_bw_signals7c.txt};
    \addplot[uclg9,only marks,mark=o,mark size=1.5pt] table[x=num,y=cap]{data/ULL-low-wp1spans80kmpnan_fw_bw_signals7c.txt};
    
    \addplot[uclg7,only marks,mark=triangle,mark size=1.5pt] table[x=num,y=cap]{data/ULL-low-wp1spans80kmp25_fw_bw_signals8c.txt};
    \addplot[uclg6,only marks,mark=triangle,mark size=1.5pt] table[x=num,y=cap]{data/ULL-low-wp1spans80kmp20_fw_bw_signals8c.txt};
    \addplot[uclg8,only marks,mark=triangle,mark size=1.5pt] table[x=num,y=cap]{data/ULL-low-wp1spans80kmp18_fw_bw_signals8c.txt};
    \addplot[uclg9,only marks,mark=triangle,mark size=1.5pt] table[x=num,y=cap]{data/ULL-low-wp1spans80kmpnan_fw_bw_signals8c.txt};

\draw[gray,densely dotted] (axis cs:0.000,485.6) -- (axis cs:0.000,511.2);
\draw[gray,densely dotted] (axis cs:0.147,493.9) -- (axis cs:0.147,511.3);
\draw[gray,densely dotted] (axis cs:0.197,501.5) -- (axis cs:0.197,511.2);
\draw[gray,densely dotted] (axis cs:0.177,504.5) -- (axis cs:0.177,511.4);
\draw[gray,densely dotted] (axis cs:0.290,505.9) -- (axis cs:0.290,511.2);
\draw[gray,densely dotted] (axis cs:0.240,499.0) -- (axis cs:0.240,511.4);
\draw[gray,densely dotted] (axis cs:0.093,491.1) -- (axis cs:0.093,511.4);

\draw[gray,densely dotted] (axis cs:0.000,475.6) -- (axis cs:0.000,449.9);
\draw[gray,densely dotted] (axis cs:0.147,483.9) -- (axis cs:0.147,462.1);
\draw[gray,densely dotted] (axis cs:0.197,491.5) -- (axis cs:0.197,472.6);
\draw[gray,densely dotted] (axis cs:0.343,498.2) -- (axis cs:0.343,482.7);
\draw[gray,densely dotted] (axis cs:0.177,494.5) -- (axis cs:0.177,475.3);
\draw[gray,densely dotted] (axis cs:0.323,501.3) -- (axis cs:0.323,486.7);
\draw[gray,densely dotted] (axis cs:0.373,505.9) -- (axis cs:0.373,491.9);
\draw[gray,densely dotted] (axis cs:0.480,508.1) -- (axis cs:0.480,497.2);
\draw[gray,densely dotted] (axis cs:0.387,504.9) -- (axis cs:0.387,498.2);
\draw[gray,densely dotted] (axis cs:0.467,507.8) -- (axis cs:0.467,494.9);
\draw[gray,densely dotted] (axis cs:0.417,503.9) -- (axis cs:0.417,492.2);
\draw[gray,densely dotted] (axis cs:0.270,497.3) -- (axis cs:0.270,484.6);
\draw[gray,densely dotted] (axis cs:0.437,502.1) -- (axis cs:0.437,489.0);
\draw[gray,densely dotted] (axis cs:0.290,495.9) -- (axis cs:0.290,479.7);
\draw[gray,densely dotted] (axis cs:0.240,489.0) -- (axis cs:0.240,469.9);
\draw[gray,densely dotted] (axis cs:0.093,481.1) -- (axis cs:0.093,458.0);

    \nextgroupplot[ytick pos=right, ylabel near ticks,
        ytick={310,340,...,490},]

    \addplot[uclg7,only marks,mark=o,mark size=1.5pt] table[x=num,y=cap]{data/ULL-low-wp3spans80kmp25_fw_bw_signals7c.txt};    
    \addplot[uclg6,only marks,mark=o,mark size=1.5pt] table[x=num,y=cap]{data/ULL-low-wp3spans80kmp20_fw_bw_signals7c.txt};
    \addplot[uclg8,only marks,mark=o,mark size=1.5pt] table[x=num,y=cap]{data/ULL-low-wp3spans80kmp18_fw_bw_signals7c.txt};
    \addplot[uclg9,only marks,mark=o,mark size=1.5pt] table[x=num,y=cap]{data/ULL-low-wp3spans80kmpnan_fw_bw_signals7c.txt};

    \addplot[uclg7,only marks,mark=triangle,mark size=1.5pt] table[x=num,y=cap]{data/ULL-low-wp3spans80kmp25_fw_bw_signals8c.txt};
    \addplot[uclg6,only marks,mark=triangle,mark size=1.5pt] table[x=num,y=cap]{data/ULL-low-wp3spans80kmp20_fw_bw_signals8c.txt};
    \addplot[uclg8,only marks,mark=triangle,mark size=1.5pt] table[x=num,y=cap]{data/ULL-low-wp3spans80kmp18_fw_bw_signals8c.txt};
    \addplot[uclg9,only marks,mark=triangle,mark size=1.5pt] table[x=num,y=cap]{data/ULL-low-wp3spans80kmpnan_fw_bw_signals8c.txt};

\draw[gray,densely dotted] (axis cs:0.000,396.1) -- (axis cs:0.000,445.5);
\draw[gray,densely dotted] (axis cs:0.147,409.6) -- (axis cs:0.147,446.4);
\draw[gray,densely dotted] (axis cs:0.197,422.9) -- (axis cs:0.197,446.7);
\draw[gray,densely dotted] (axis cs:0.343,433.2) -- (axis cs:0.343,446.4);
\draw[gray,densely dotted] (axis cs:0.177,427.3) -- (axis cs:0.177,448.1);
\draw[gray,densely dotted] (axis cs:0.323,435.7) -- (axis cs:0.323,447.1);
\draw[gray,densely dotted] (axis cs:0.417,441.8) -- (axis cs:0.417,446.9);
\draw[gray,densely dotted] (axis cs:0.270,433.3) -- (axis cs:0.270,447.1);
\draw[gray,densely dotted] (axis cs:0.437,437.8) -- (axis cs:0.437,446.4);
\draw[gray,densely dotted] (axis cs:0.290,430.3) -- (axis cs:0.290,446.9);
\draw[gray,densely dotted] (axis cs:0.240,418.3) -- (axis cs:0.240,447.0);
\draw[gray,densely dotted] (axis cs:0.093,405.5) -- (axis cs:0.093,446.7);

\draw[gray,densely dotted] (axis cs:0.000,386.1) -- (axis cs:0.000,350.8);
\draw[gray,densely dotted] (axis cs:0.147,399.6) -- (axis cs:0.147,368.9);
\draw[gray,densely dotted] (axis cs:0.197,412.9) -- (axis cs:0.197,383.3);
\draw[gray,densely dotted] (axis cs:0.343,423.2) -- (axis cs:0.343,396.7);
\draw[gray,densely dotted] (axis cs:0.177,417.3) -- (axis cs:0.177,388.7);
\draw[gray,densely dotted] (axis cs:0.323,425.7) -- (axis cs:0.323,400.9);
\draw[gray,densely dotted] (axis cs:0.373,432.5) -- (axis cs:0.373,405.7);
\draw[gray,densely dotted] (axis cs:0.480,433.0) -- (axis cs:0.480,411.7);
\draw[gray,densely dotted] (axis cs:0.387,433.6) -- (axis cs:0.387,412.7);
\draw[gray,densely dotted] (axis cs:0.467,436.1) -- (axis cs:0.467,409.4);
\draw[gray,densely dotted] (axis cs:0.417,431.8) -- (axis cs:0.417,405.1);
\draw[gray,densely dotted] (axis cs:0.270,423.3) -- (axis cs:0.270,396.2);
\draw[gray,densely dotted] (axis cs:0.437,427.8) -- (axis cs:0.437,404.9);
\draw[gray,densely dotted] (axis cs:0.290,420.3) -- (axis cs:0.290,396.1);
\draw[gray,densely dotted] (axis cs:0.240,408.3) -- (axis cs:0.240,381.0);
\draw[gray,densely dotted] (axis cs:0.093,395.5) -- (axis cs:0.093,363.3);

    \end{groupplot}
    
    \draw (power_graph c1r1.north west) node[xshift=1em,yshift=-.8em] {\small (a)};
    \draw (power_graph c2r1.north west) node[xshift=1em,yshift=-.8em] {\small (b)};
    \draw (power_graph c1r2.north west) node[xshift=1em,yshift=-.8em] {\small (c)};
    \draw (power_graph c2r2.north west) node[xshift=1em,yshift=-.8em] {\small (d)};
    
\end{tikzpicture}

%% file: snr.tex
\pgfplotstableread{data/ndff_snr.txt}\datandff
\pgfplotstableread{data/ull_snr.txt}\dataull

\vspace{-2em}
\begin{tikzpicture}
\renewcommand{\THz}[1]{\fpeval{299792.458/(#1)}}

\begin{groupplot}[
    group style={
        group name=snr_graph,
        group size=1 by 4,
        x descriptions at=edge bottom,
        vertical sep=0.2em,
        horizontal sep=0.2em,
    },
    width=1.05\linewidth, height=3.4cm,
    unbounded coords=jump,
    ymajorgrids=true,
    xmajorgrids=true,
    x dir=reverse,
    xlabel={Wavelength (${\rm nm}$)},
    xlabel near ticks,
    xmin=\THz{1650},
    xmax=\THz{1260},
    ylabel near ticks,
    ylabel shift = -2 pt,
    xlabel shift = -2 pt,
    ymin=-2,ymax=19.5,
    xticklabels={$1260$,$1360$,$1460$,$1560$,$1620$},
    xtick={\THz{1260},\THz{1360},\THz{1460},\THz{1560},\THz{1620}},
    yticklabel style={/pgf/number format/.cd,fixed,fixed zerofill,precision=0}
]

\nextgroupplot[]

\node[anchor=mid] at (axis cs:\THz{1310},0.5){O};
\node[anchor=mid] at (axis cs:\THz{1430},0.5){E};
\node[anchor=mid] at (axis cs:\THz{1495},0.5){S};
\node[anchor=mid] at (axis cs:\THz{1547.5},0.5){C};
\node[anchor=mid] at (axis cs:\THz{1595},0.5){L};

\addplot[Set1-D,only marks, mark=*, mark size=.5pt] table[x=freq,y=k1_s1_p18] \datandff {};
\addplot[Set1-E,only marks, mark=*, mark size=.5pt] table[x=freq,y=k1_s1_p18_o] \datandff {};

\addplot[Set1-C,only marks, mark=*, mark size=.5pt] table[x=freq,y=k1_s1_p18] \dataull {};
\addplot[Set1-B,only marks, mark=*, mark size=.5pt] table[x=freq,y=k1_s1_p18_o] \dataull {};

\nextgroupplot[]

\addplot[Set1-D,only marks, mark=*, mark size=.5pt] table[x=freq,y=k1_s3_p18] \datandff {};
\addplot[Set1-E,only marks, mark=*, mark size=.5pt] table[x=freq,y=k1_s3_p18_o] \datandff {};

\addplot[Set1-C,only marks, mark=*, mark size=.5pt] table[x=freq,y=k1_s3_p18] \dataull {};
\addplot[Set1-B,only marks, mark=*, mark size=.5pt] table[x=freq,y=k1_s3_p18_o] \dataull {};

\nextgroupplot[
    ylabel={SNR (dB)}, ylabel style={xshift=2em},
    legend style={font=\footnotesize,at={(rel axis cs:1.02,0.05)},legend columns=2,anchor=south east},
    legend entries={Fibre-A $^{(1)}$, Fibre-A $^{(2)}$, Fibre-B $^{(1)}$, Fibre-B $^{(2)}$}
]

\addlegendimage{Set1-D,ultra thick}
\addlegendimage{Set1-E,ultra thick}
\addlegendimage{Set1-C,ultra thick}
\addlegendimage{Set1-B,ultra thick}

\addplot[Set1-D,only marks, mark=*, mark size=.5pt] table[x=freq,y=k1_s1_p20] \datandff {};
\addplot[Set1-E,only marks, mark=*, mark size=.5pt] table[x=freq,y=k1_s1_p20_o] \datandff {};

\addplot[Set1-C,only marks, mark=*, mark size=.5pt] table[x=freq,y=k1_s1_p20] \dataull {};
\addplot[Set1-B,only marks, mark=*, mark size=.5pt] table[x=freq,y=k1_s1_p20_o] \dataull {};

\nextgroupplot[]

\addplot[Set1-D,only marks, mark=*, mark size=.5pt] table[x=freq,y=k1_s3_p20] \datandff {};
\addplot[Set1-E,only marks, mark=*, mark size=.5pt] table[x=freq,y=k1_s3_p20_o] \datandff {};

\addplot[Set1-C,only marks, mark=*, mark size=.5pt] table[x=freq,y=k1_s3_p20] \dataull {};
\addplot[Set1-B,only marks, mark=*, mark size=.5pt] table[x=freq,y=k1_s3_p20_o] \dataull {};

\end{groupplot}

\draw (snr_graph c1r1.north west) node[xshift=-.2em,yshift=-.5em,right] {\footnotesize (a) 1span 18dB};
\draw (snr_graph c1r2.north west) node[xshift=-.2em,yshift=-.5em,right] {\footnotesize (b) 3span 18dBm};
\draw (snr_graph c1r3.north west) node[xshift=-.2em,yshift=-.5em,right] {\footnotesize (c) 1span 20dBm};
\draw (snr_graph c1r4.north west) node[xshift=-.2em,yshift=-.5em,right] {\footnotesize (d) 3span 20dBm};

\end{tikzpicture}